\documentclass[12pt,preprint2]{emulateapj}
\usepackage{pslatex}
\usepackage[T1]{fontenc}
\usepackage[latin1]{inputenc}
\usepackage{subfigure}
\usepackage{amsmath}
\usepackage{graphicx,graphics}
\usepackage{amssymb}
\usepackage[english]{babel}

\newcommand{\gap}{\;\rlap{\lower 2.5pt \hbox{$\sim$}}\raise 1.5pt\hbox{$>$}\;}
\newcommand{\lap}{\;\rlap{\lower 2.5pt \hbox{$\sim$}}\raise 1.5pt\hbox{$<$}\;}
\newcommand{\beq}{\begin{equation}}
\newcommand{\eeq}{\end{equation}}

\shorttitle{Mass Deficits in Galaxies}
\shortauthors{David Merritt}
\begin{document}

\title{Mass Deficits, Stalling Radii, and the Merger Histories of Elliptical Galaxies}

\author{David Merritt}
\affil{Department of Physics, Rochester Institute of Technology,
Rochester, NY 14623}

\begin{abstract}
A binary supermassive black hole leaves an imprint
on a galactic nucleus in the form of a ``mass deficit,''
a decrease in the mass of the nucleus due to ejection of
stars by the binary.
The magnitude of the mass deficit is in principle related to
the galaxy's merger history, but the relation has never
been quantified.
Here, high-accuracy $N$-body simulations are used to
calibrate this relation.
Mass deficits are shown to be 
$M_{def}\approx 0.5M_{12}$, with $M_{12}$ 
the total mass of the binary;
the coefficient in this relation
depends only weakly on $M_2/M_1$ or on the galaxy's
pre-existing density profile.
Hence, after ${\cal N}$ mergers, 
$M_{def}\approx 0.5{\cal N}M_\bullet$
with $M_\bullet$ the final (current) black hole mass.
When compared with observed mass deficits, this result
implies $1\lap {\cal N}\lap 3$,
in accord with hierarchical galaxy formation models.
Implications for binary stalling radii,
the origin of hyper-velocity stars,
and the distribution of dark matter at the centers of galaxies
are discussed.
\end{abstract}

\section{Introduction}

Galaxy mergers bring supermassive black holes (SBHs)
together \citep{bbr-80}, and binary SBHs are 
increasingly invoked to explain the properties
of normal and active galaxies \citep{komossa-03}, 
including AGN variability \citep{valtaoja-00,xie-03},
the  bending and precession of radio jets \citep{roos-93,romero-00},
$X$- and $Z$-shaped radio lobes \citep{me-02,gopal-03}
and the presence of cores in bright elliptical galaxies
\citep{mm-02,graham-04}.
Binary SBHs may also be responsible for the 
high-velocity stars observed in the halo of the
Milky Way \citep{yu-03,haardt-06}, 
and for other populations that appear to 
have been ejected from galaxies, including
intra-cluster planetary nebulae \citep{hb-05}.

The dynamical interaction of a massive binary
with stars in a galactic nucleus is often
discussed with reference to rate equations
derived from scattering experiments
\citep{hills-83,mv-92,quinlan-96}.
A massive binary hardens at a rate
\beq
{d\over dt}\left({1\over a}\right) = H {G\rho\over\sigma}
\label{eq:h}
\eeq
where $a$ is the binary semi-major axis, $\rho$ and $\sigma$
are the stellar density and velocity dispersion, and
$H$ is a dimensionless rate coefficient that depends on the
binary mass ratio, eccentricity, and hardness.
The mass in stars ejected by the binary satisfies
\beq
{dM_{ej}\over d\ln(1/a)} = JM_{12},
\label{eq:mej}
\eeq
where $M_{12}\equiv M_1+M_2$, the binary mass,
and $J$ is a second dimensionless coefficient.
Equations~(\ref{eq:h}) and~(\ref{eq:mej})
have been used to estimate
evolution rates of binary SBHs in galactic
nuclei (e.g. Volonteri et al. 2003a, b; Sesana et al. 2004) and
to compute the mass in stars ejected from
the galaxy by the binary (e.g. Holey-Bockelmann et al. 2005;
Haardt et al. 2006).
These equations are also the basis for many hybrid schemes
which imbed a binary into a model of the
host galaxy \citep{zier-01,yu-02,wang-05}.

Equations~(\ref{eq:h}) and~(\ref{eq:mej}) are not 
particularly useful however when computing the 
binary-induced change in the distribution of stars 
in a nucleus,
which is the topic of the current paper.
$N$-body experiments show that the two SBHs
substantially change the stellar density in a short time,
of order the galaxy crossing time or less,
{\it before} they form a tightly-bound pair
(e.g. Figure 4 of Milosavljevic \& Merritt 2001).
In other words, by the time that equations like 
(\ref{eq:h}) and (\ref{eq:mej}) start to become valid,
a considerable change has already taken
place in the stellar distribution.
Furthermore, in many galaxies,
evolution of the binary would be expected to
stall at about the same time that it
becomes hard \citep{valtonen-96}.
Even if the mass ejected by the binary before this
time could
be accurately computed from an equation like~(\ref{eq:mej}),
it would still be difficult to relate $M_{ej}$ to observable changes
in the nuclear density profile.

$N$-body techniques would seem to be the solution
to these problems;
but $N$-body integrations are plagued by spurious relaxation
and other discreteness effects,
which cause an embedded binary to continue
evolving even in circumstances where a real
binary would stall \citep{bms-05}.
The counterpart of these discreteness effects in real
galaxies is two-body relaxation,
but -- especially in the luminous galaxies that show 
evidence of the ``scouring'' effects of a massive binary --
central relaxation times are much too long to
significantly affect the supply of stars to the binary \citep{yu-02}.
$N$-body simulations that are dominated by discreteness
effects can not be scaled to the essentially
collisionless regime of real galaxies.

However, there has been much progress in the
art of $N$-body simulation in recent years,
such that high-accuracy, direct-summation
integrations are now feasible with particle 
numbers as large as $10^6$ \citep{dorband-03,gzt-04,fmk-05,bms-05}.
Given such high values of $N$, it becomes
possible to separate the rapid, early
phases of binary evolution -- which,
when appropriately normalized, should
be independent of $N$ -- from the late,
slow stages that are driven
by $N$-dependent process like collisional
loss-cone repopulation.

This is the approach adopted in the present paper.
$N$-body simulations are used to follow the
evolution of a galaxy containing a central
massive object and a second, inspiralling 
point mass.
The integrations are continued until the two 
massive particles form a tight binary.
The time at which the evolution of the binary
switches from rapid -- i.e. $N$-independent -- to 
slow -- i.e. $N$-dependent -- is identified,
and the properties of the galaxy are recorded at that time.
The degree to which a clear separation of the two regimes is possible,
for a given $N$, depends on the binary mass ratio,
becoming more difficult as the mass ratio becomes
more extreme.
Results are presented here for mass ratios in the range
$0.025 \le M_2/M_1 \le 0.5$.

We find (\S4) that the {\it mass deficit} --
the difference in integrated mass between initial and 
final nuclear density profiles --
at the end of the rapid evolutionary phase
is proportional to the total mass of the binary,
$M_{def}\approx 0.5 M_{12}$, with only a weak
dependence on $M_2/M_1$ or on the initial density
profile.
This result can be motivated by simple arguments (\S2):
the smaller $M_2$, the tighter the binary which it forms
before stalling.

A mass deficit of $\sim 0.5M_\bullet$
is at least a factor two smaller than the 
typical mass deficits observed in bright
elliptical galaxies
\citep{mm-02,graham-04,acs-6}
but we show via an additional set of $N$-body
simulations (\S 5) that the effect of
binary SBHs on a nucleus is {\it cumulative}, 
scaling roughly in proportion both to the number of 
mergers and to the final mass of the SBH.
Hence, observed values of $M_{def}/M_\bullet$ can be
used to constrain the number ${\cal N}$ of mergers that have
taken place since the era at which the SBHs first formed.
We find (\S 6) that $1\lap {\cal N}\lap 3$,
consistent with expectations from hierarchical structure
formation theory.
Finally, in \S 7 and \S 8, the implications for
binary stalling radii, ejection of high-velocity stars,
and the distribution of dark matter are discussed.

\section{Stages of Binary Evolution}

Here we review the stages of binary SBH evolution
and discuss the connection between evolution of
a binary in $N$-body simulations and in real galaxies.

Let $M_1$ and $M_2$ to be the masses of the two
components of the binary, and write $M_{12}\equiv M_1+M_2$, 
$q\equiv M_2/M_1\le 1$, and $\mu\equiv M_1M_2/M_{12}$.
In what follows we assume that the larger SBH is located 
initially at the center of the galaxy and that the smaller
SBH spirals in.

The evolution of the massive binary is customarily divided
into three phases.

\noindent 1. At early times, the orbit of the smaller SBH
decays due to dynamical friction from the stars.
This phase ends when the separation $R_{12}$ between
the two SBHs is $\sim r_h$, the gravitational influence radius 
of the larger hole.
We define $r_h$ in the usual way as the radius of
a sphere around $M_1$ that encloses a stellar mass of $2M_1$:
\beq
M_\star(r_h) = 2M_1.
\label{eq:rh}
\eeq

\noindent 2. When $R_{12}$ falls below $\sim r_h$, 
the two SBHs form a bound pair.
The separation between the two SBHs drops rapidly 
in this phase, due both to dynamical friction 
acting on $M_2$, and later to ejection of stars
by the binary \citep{mm-01}.
The motion of the smaller SBH around the larger
is approximately Keplerian in this phase;
we denote the semi-major axis by $a$ and the 
eccentricity by $e$.
The relative velocity of the two SBHs 
for $e=0$ is
\beq
V_{bin}=\sqrt{GM_{12}\over a} 
\eeq
and the binary's energy is
\beq
E = -{GM_1M_2\over 2a}.
\eeq

\noindent 3. The rapid phase of binary
evolution comes to an end when the binary's
binding energy reaches 
$\sim M_{12}\sigma^2$, i.e. when 
$a\approx a_h$, where $a_h$ is the semi-major axis of a ``hard''
binary, sometimes defined as
\beq
a_h = {G\mu\over 4\sigma^2}.
\label{eq:ah}
\eeq
If one writes $r_h=GM_1/\sigma^2$ -- 
equivalent to the definition (\ref{eq:rh}) 
in a nucleus with $\rho\sim r^{-2}$ -- then
$a_h$ can also be written
\beq
a_h = {q\over (1+q)^2} {r_h\over 4}.
\label{eq:ahapprox}
\eeq

The exact definition of a ``hard'' binary varies
from author to author, and this vagueness 
reflects the difficulty of relating the evolution of
an isolated binary to one embedded in a galaxy.
An isolated binary in a fixed background begins
to harden at an approximately constant rate,
$(d/dt)(1/a)\approx {\rm const.}$, when
$a\lap a_h$ \citep{quinlan-96}.
But in a real galaxy, a hard binary would 
efficiently eject all stars on intersecting orbits,
and its hardening rate would suddenly {\it drop}
at $a\approx a_h$.
This effect has been observed in $N$-body experiments 
with sufficiently large $N$ \citep{mf-04,bms-05}.

Supposing that the binary ``stalls'' at $a\approx a_h$,
it will have given up an energy
\begin{subequations}
\begin{eqnarray}
\Delta E &\approx& -{GM_1M_2\over 2r_h} + {GM_1M_2\over 2a_h} \\
&\approx& -{1\over 2}M_2\sigma^2 + 2M_{12}\sigma^2 \\
&\approx& 2 M_{12}\sigma^2
\end{eqnarray}
\end{subequations}
to the stars in the nucleus.
In other words, the energy transferred from the binary to 
the stars is roughly proportional to the 
{\it combined} mass of the two SBHs.
This result suggests that the binary will displace a mass
in stars of order its own  mass, independent
of the mass of the infalling hole.
This prediction is verified in the $N$-body
simulations presented below.

Since the ``hard'' binary separation $a_h$ is 
ill-defined for binaries embedded in galaxies,
we define here a more useful quantity.
The {\it stalling radius} $a_{stall}$ is defined as the 
separation at which evolution of the binary would halt,
$d(1/a)/dt=0$, in the absence of any mechanism to
re-supply the stellar orbits.
For instance, if the galaxy potential were completely 
smooth, the binary's decay would  halt once all the stars
on orbits intersecting the binary had been ejected
or otherwise (e.g. due to shrinkage of the binary)
stopped interacting with it.
The motivation for this definition is the very
long time scales, in luminous elliptical galaxies, 
for orbital re-population by two-body encounters.
Even in galaxies with relatively short
central relaxation times,
the binary's hardening rate would 
drop drastically at $a=a_{stall}$ and the likelihood 
of finding the binary at a separation near $a_{stall}$
would be high.

This definition implies a dependence of $a_{stall}$
on the mass ratio of the binary, but (unlike 
Equation~\ref{eq:ah}) it also implies a dependence on 
the initial density profile and shape (spherical,
axisymmetric, triaxial) of the galaxy, since these
factors may influence the mass in stars that can
interact with the binary.
The definition is operational:
it can only be applied by ``doing the experiment,''
i.e. imbedding the binary in a galaxy model, turning off
gravitational perturbations (aside from those due
to the binary itself), and observing when the decay halts.

In a numerical simulation with finite $N$, 
gravitational encounters will continue 
supplying stars to the binary at rates much higher
than those in real galaxies.
One can still estimate $a_{stall}$ if the
particle number is large enough that a clear change
takes place in the hardening rate at some value of
$a$.
Better still, by repeating the experiment with different
values of $N$, one can
hope to show that the rapid evolutionary phase comes
to an end at a well-defined time and that subsequent evolution occurs at
a rate that is a decreasing function of $N$.

One advantage of defining $a_{stall}$ in this operational
way is that it eliminates the need for many of the
qualitative distinctions that have been made in the past
between the different regimes of binary evolution, e.g. 
``soft binaries'' vs. ``hard binaries,'' hardening 
via ``dynamical friction'' vs. hardening via
``loss-cone draining,'' etc. \citep{yu-02}.
It also allows for mechanisms that are not reproduced
in the scattering experiments on which Equations (1) and (2)
are based,
e.g. the ``secondary  slingshot'' \citep{mm-03},
or the effect of the changing nuclear density on
the rate of supply of stars to the binary.

It would seem natural to compute $a_{stall}$ 
using so-called ``collisionless'' $N$-body codes
that approximate the gravitational potential
via smooth basis functions \citep{cb-73,albada-77,ho-92}.
Such algorithms have in fact been applied to the binary SBH
problem \citep{quinlan-97,hss-02,chl-03} but with results that
are not always consistent with those of direct-summation
codes or with the predictions of loss-cone theory 
(compare Chatterjee, Henquist \& Loeb 2003 with 
Berczik, Merritt \& Spurzem 2005).
These inconsistencies may be due to difficulties associated 
with incorporating a binary into a collisionless code,
ambiguities about the best choice of origin for the potential expansion,
etc.
Until these issues can be resolved, direct-summation $N$-body
codes seem a safer choice.

\begin{table}
\begin{center}
\caption{Parameters of the $N$-body integrations \label{tbl-1}}
\begin{tabular}{ccccccc}
\tableline\tableline
$\gamma$ & $M_2/M_1$ & $r_h$    & $r_h'$  & $a_{\rm stall}$      & $a_{\rm stall}/r_h'$ & $M_{\rm def}/M_{12}$\\
\tableline
$0.5$    & $0.5$     & $0.264$  & $0.37$  & $1.49\times 10^{-2}$ & $4.08\times 10^{-2}$ & $0.60$  \\ 
$0.5$    & $0.25$    & $0.264$  & $0.32$  & $1.18\times 10^{-2}$ & $3.69\times 10^{-2}$ & $0.46$  \\
$0.5$    & $0.1$     & $0.264$  & $0.30$  & $5.38\times 10^{-3}$ & $1.82\times 10^{-2}$ & $0.33$ \\
$0.5$    & $0.05$    & $0.264$  & $0.29$  & $3.60\times 10^{-3}$ & $1.26\times 10^{-2}$ & $0.27$ \\
$0.5$    & $0.025$   & $0.264$  & $0.28$  & $1.91\times 10^{-3}$ & $6.82\times 10^{-3}$ & $0.24$ \\
&&&&& \\
$1.0$    & $0.5$     & $0.165$  & $0.24$  & $7.75\times 10^{-3}$ & $3.23\times 10^{-2}$ & $0.62$      \\
$1.0$    & $0.25$    & $0.165$  & $0.22$  & $4.83\times 10^{-3}$ & $2.25\times 10^{-2}$ & $0.54$     \\
$1.0$    & $0.1$     & $0.165$  & $0.19$  & $2.37\times 10^{-3}$ & $1.23\times 10^{-2}$ & $0.45$     \\
$1.0$    & $0.05$    & $0.165$  & $0.19$  & $1.36\times 10^{-3}$ & $7.27\times 10^{-3}$ & $0.39$     \\
$1.0$    & $0.025$   & $0.165$  & $0.18$  & $6.21\times 10^{-4}$ & $3.36\times 10^{-3}$ & $0.34$     \\
&&&&& \\
$1.5$    & $0.5$     & $0.0795$ & $0.13$  & $3.66\times 10^{-3}$ & $2.77\times 10^{-2}$ & $0.63$     \\
$1.5$    & $0.25$    & $0.0795$ & $0.11$  & $2.18\times 10^{-3}$ & $1.95\times 10^{-2}$ & $0.56$     \\
$1.5$    & $0.1$     & $0.0795$ & $0.10$  & $1.26\times 10^{-3}$ & $1.29\times 10^{-2}$ & $0.41$     \\
$1.5$    & $0.05$    & $0.0795$ & $0.10$  & $7.78\times 10^{-4}$ & $8.19\times 10^{-3}$ & $0.38$     \\
$1.5$    & $0.025$   & $0.0795$ & $0.09$  & $4.60\times 10^{-4}$ & $4.84\times 10^{-3}$ & $0.39$     \\
\tableline
\end{tabular}
\end{center}
\end{table}

\section{Models and Methods}

The galaxy models and $N$-body techniques used here
are similar to those described in \cite{szell-05}
(hereafter Paper I).
In brief, Monte-Carlo realizations of steady-state,
spherical galaxy models were constructed using Dehnen's 
(1993) density law.
The models contained an additional, central point mass 
representing the more massive of the two SBHs.
The mass of this particle, $M_1$, was always set to $0.01$
in units where the total mass in stars $M_{gal}$ was one.
(The Dehnen scale length $r_D$ and the gravitational constant 
$G$ are also set to unity in what follows.)
Stellar positions and velocities were generated from the
unique isotropic phase-space distribution function
that reproduces Dehnen's $\rho(r)$ in the combined gravitational
potential of the stars and the central point mass.
The second SBH was introduced into this model at $t=0$,
and given a velocity roughly $1/2$ of the circular orbital
velocity.
In the models with $\gamma=0.5$ and $\gamma=1.0$,
the initial separation of the two massive particles 
was $1.6$, while in the models with $\gamma=1.5$
the initial separation was $0.5$.
Each realization was then integrated forward using the $N$-body
integrator described in \cite{msm-05}, an adaptation
of {\tt NBODY1} \cite{Aarseth:99} to the GRAPE-6 special-purpose
computer.
Close encounters between the two SBH particles, 
and between the SBHs and stars,
were regularized using the algorithm 
of Mikkola and Aarseth \citep{MA:90,MA:93}.

Three parameters suffice to define the initial
models: the binary mass ratio $q$, the central density
slope of the Dehnen model $\gamma\equiv - d\log\rho/d\log r$,
and the number of particles $N$.
Five values of the binary mass ratio were considered:
$q=(0.5,0.25,0.1,0.05,0.025)$.
For each choice of $q$, three values of $\gamma$ were
used: $0.5, 1.0$, and $1.5$.
Finally, each of these 15 initial models was integrated with
two different values of $N$: $1.2\times 10^5$ and $2.0\times 10^5$.
Mass deficits were computed as described in Paper I.

The parameters of the $N$-body integrations are summarized in Table 1.
Columns three and four give two different estimates
of the binary's influence radius.
The first, $r_h$, is the radius containing a mass in stars 
equal  to twice $M_1$ in the initial model.
The second is the radius containing a mass in
stars equal to twice $M_1+M_2$ at a time $t=t_{stall}$,
the estimated stalling time.
The second definition, $r_h'$, is the relevant
one when comparing the results of the $N$-body
integrations to a real galaxy which has already
experienced the scouring effects of a binary SBH.
The other columns in Table 1 give estimates of
the stalling radius $a_{stall}\equiv a(t_{stall})$
and the mass deficit $M_{def}(t_{stall})$,
derived as described below.

\begin{figure}
\centering
\includegraphics[scale=0.45,angle=-90.]{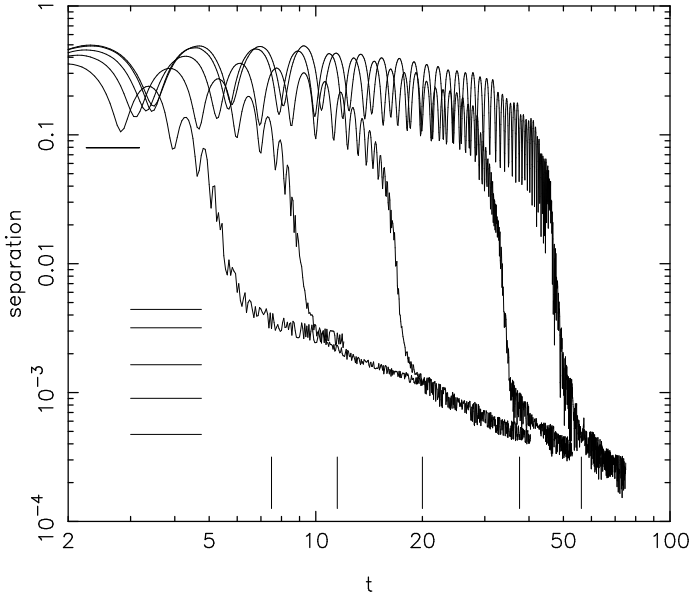}
\caption{Evolution of the binary separation in five
$N$-body integrations with $\gamma=1.5$ and $N=200$K;
binary mass ratios are, from left to right,
$0.5,0.25,0.1,0.05,0.025$.
Vertical lines show the times identified as $t=t_{stall}$.
The upper horizontal line indicates $r_h$,
the influence radius of the more massive hole in the
initial model.
Lower horizontal lines show $a_h$ as defined in 
Equation~(\ref{eq:ahapprox}).
The rapid phase of decay  continues until 
$a\approx a_h$,
with the result that the binary's binding energy 
at the end of this phase is nearly independent of $M_2$.
\label{fig:rbinoft}
}
\end{figure}

\section{Results}

Figure~\ref{fig:rbinoft} illustrates the
evolution of the relative orbit in the five integrations
with $\gamma=1.5$ and $N=200K$.
The three phases discussed above are evident.
(1) The separation $R_{12}$ between the two massive particles
gradually decreases as dynamical
friction extracts angular momentum from the orbit of $M_2$.
(2) When $R_{12}\approx r_h$, the rapid phase of
binary evolution begins.
(3) When the separation drops to $\sim a_h$,
evolution again slows.
Figure~\ref{fig:rbinoft} shows clearly that the
separation of the two massive particles
at the start of the final phase is
smaller for smaller $M_2$.

\begin{figure}
\centering
\includegraphics[scale=0.50,angle=0.]{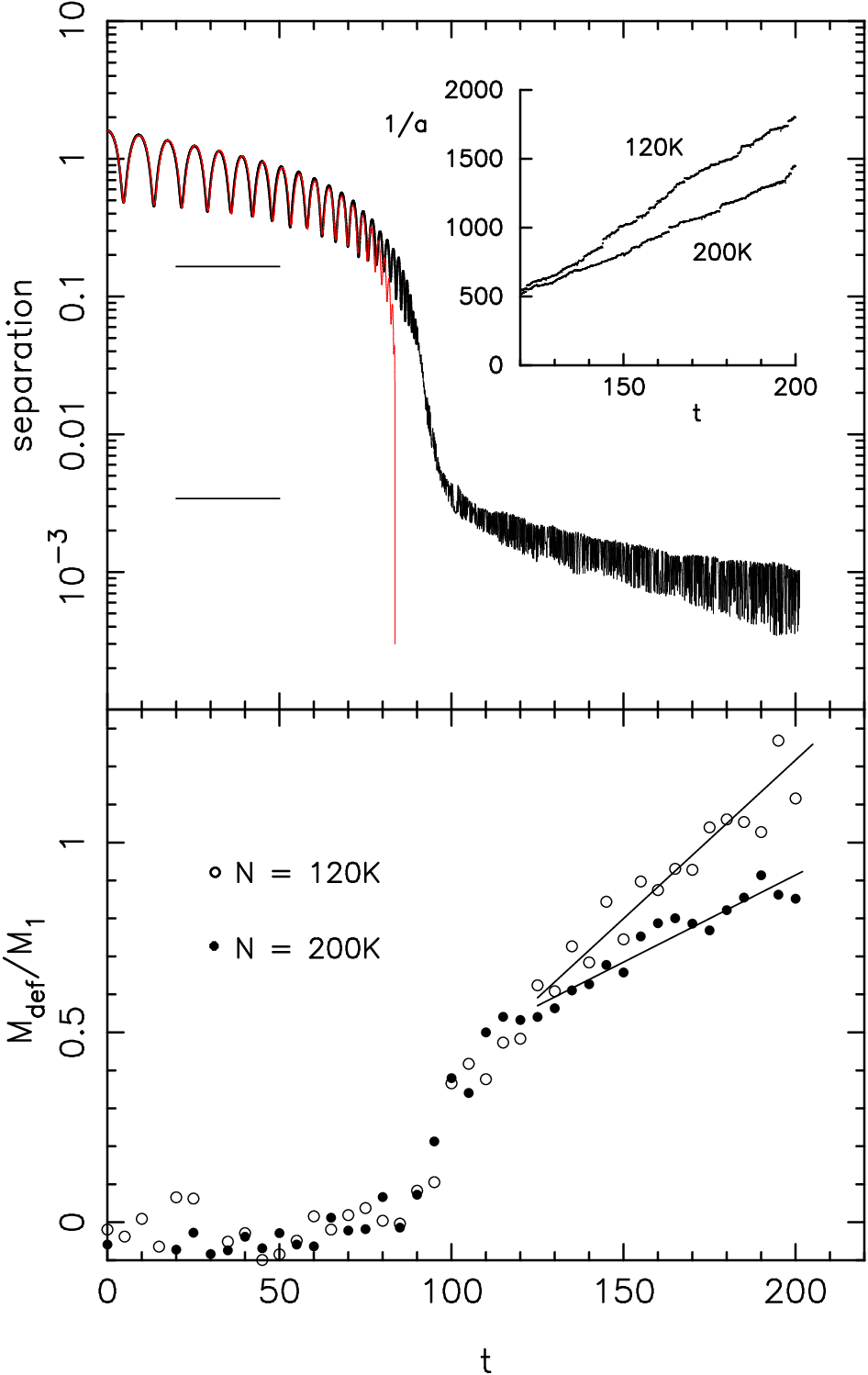}
\caption{{\it Upper panel:} Thick (black) line shows evolution 
of the binary separation in the $N$-body integration with 
$\gamma=1.0$, $M_2/M_1=0.1$, $N=200$K.
Thin (red) line is the evolution predicted by the
dynamical friction equation (\ref{eq:chandra})
assuming a fixed galaxy.
Horizontal lines indicate $r_h$ and $a_h$,
the latter as defined in Equation~(\ref{eq:ahapprox}).
The inset shows the evolution of the inverse semi-major
axis of the binary for this integration, 
and for a second integration with $N=120$K.
{\it Lower panel:} The mass deficit, as defined in the
text, for the same two $N$-body integrations.
Lines show least-squares fits to the time intervals 
$80\le t\le 110$ and $110\le t \le 200$.
\label{fig:binary}
}
\end{figure}

Figure~\ref{fig:rbinoft} suggests that the stalling
radius is of order $a_h$, but better 
estimates of $a_{stall}$ can be made
by comparing integrations of the same model 
carried out with different $N$.
Figure~\ref{fig:binary} makes the comparison for the model 
with $\gamma=1.0$ and $q=0.1$.
After $\sim 15$ orbits, the mean separation between the
two massive particles has dropped from
$\sim 1$ to $\sim 0.2\approx r_h$
and the second, rapid phase of binary hardening begins.
Prior to this time, the two integrations with different
$N$ find essentially identical evolution of $R_{12}$.
As discussed above, the evolution during this phase is
due to some combination of dynamical friction acting
on $M_2$ and ejection of stars by the increasingly-hard
binary, both of which are $N$-independent processes.
The contribution of dynamical friction to the evolution
can be estimated using Chandrasekhar's (1943) expression
for the dynamical friction force:
\begin{subequations}
\begin{eqnarray}
\langle \Delta {\rm v}_\parallel\rangle &=& 
-4\pi(4\pi G^2 M_2\rho) \int_0^{\infty}d{\rm v}_f \left({{\rm v}_f\over {\rm v}}\right)^2 f_f({\rm v}_f) H_1({\rm v},{\rm v}_f), \\
H_1 &=& \left\{ \begin{array}{ll}
	\ln\Lambda & \mbox{if ${\rm v}>{\rm v}_f$,} \\
	0 & \mbox{if ${\rm v}<{\rm v}_f$}.
	\end{array}
	\right.
	\label{twoh}
\end{eqnarray}
\label{eq:chandra}
\end{subequations}
This is the standard approximation in which $\ln\Lambda$
is ``taken out of the integral''; ${\rm v}$ is the velocity of the
massive object, ${\rm v}_f$ is the velocity of a field star,
and $f_f({\rm v}_f)$ is the field-star velocity distribution,
normalized to unit total number and assumed fixed in time.
The result of integrating equation (\ref{eq:chandra}),
with $\ln\Lambda = 5.7$,
is shown in the upper panel of Figure~\ref{fig:binary} as the red (thin) line.
Chandrasekhar's formula accurately reproduces the
evolution of the relative orbit until a time $t\approx 80$,
after which it predicts that the separation
should drop to zero at a finite time.
However as shown in the lower panel of Figure~\ref{fig:binary},
by $t=80$ -- roughly at the start of the second 
evolutionary phase -- the infall of $M_2$ has begun to displace stars and
lower the central density, and the dynamical
friction force must drop below the value predicted by equation
(\ref{eq:chandra}) with fixed $\rho$ and $f_f$.

The inset in Figure~\ref{fig:binary} 
 shows the inverse semi-major axis of the binary,
$1/a$, versus time in the two integrations of this 
model with different $N$.
The two curves are coincident until 
$a\approx a_h$ but diverge at later times.
This is the expected result \citep{bms-05}:
once the hard binary has ejected most of the stars
on intersecting orbits, continued hardening
requires orbital repopulation which occurs on
a time scale that is roughly proportional to $N$.
The difference between the two integrations is also
apparent in the lower panel of Figure~\ref{fig:binary},
which shows mass deficits in the two integrations.
Based on this comparison, the stalling radius -- the
value of $a$ at which the evolution ceases to be independent
of particle number -- is $a_{stall}=a(t\approx 110)\approx 2\times 10^{-3}$,
and the mass deficit when $a=a_{stall}$ is 
$M_{def}\approx 0.5 M_1$.

In order to make still more accurate estimates of
$t_{stall}$ and $a_{stall}$, the hardening rate,
\beq
s(t) \equiv {d\over dt} \left({1\over a}\right),
\eeq
was estimated from the $N$-body data by fitting
smoothing splines to the measured values of
$a^{-1}$ vs. $t$ and differentiating.
In a collisionless galaxy, $s(t)$ would reach a peak
value during the rapid phase of binary evolution,
then drop rapidly to zero as the stars on intersecting orbits
are ejected by the binary; $t_{stall}$ would be the
time at which $s\approx 0$.
In the $N$-body integrations, $s$ will never
fall completely to zero since gravitational
encounters continue to scatter stars into
the binary's loss cone.

\begin{figure}
\centering
\includegraphics[scale=0.4,angle=-90.]{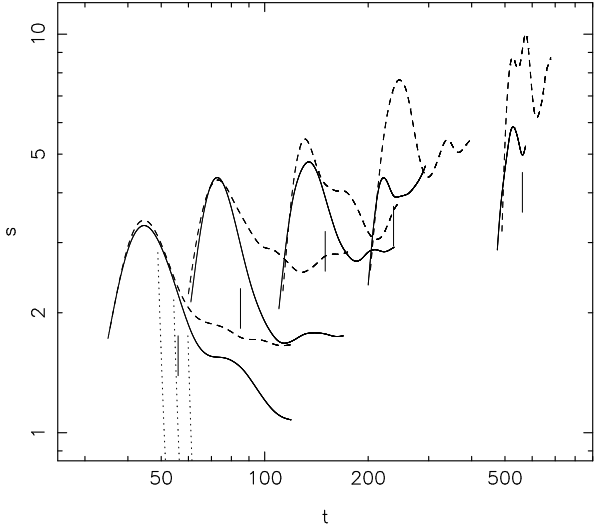}
\caption{Evolution of the binary hardening rate
$s\equiv (d/dt)(1/a)$ in 
$N$-body integrations with $\gamma=0.5$;
binary mass ratios are, from left to right,
 $0.5,0.25,0.1,0.05,0.025$.
Solid curves are for $N=200$K and dashed curves 
are for $N=120$K.
Vertical solid lines are the estimates of $t_{stall}$.
Nearly-vertical dashed lines show the approximate,
asymptotic behavior of $s(t)$ in a purely collisionless galaxy,
assuming $t(a_h) = (40,50,60)$.
\label{fig:s}
}
\end{figure}

Figure~\ref{fig:s} shows $s(t)$ as extracted from
the $N$-body integrations with $\gamma=0.5$.
In each case, $s$ reaches a peak value during the
rapid phase of evolution and then declines.
In the models with largest $M_2$, loss-cone
repopulation is least efficient, and 
the hardening rates in the two integrations
with different $N$ ``track'' each other well past the peak.
As $M_2$ is decreased, loss cone repopulation
becomes more efficient and the two $s(t)$ curves
deviate from each other at progressively earlier times with
respect to the peak.

Even in the absence of loss cone repopulation,
a binary in a collisionless galaxy would continue to 
evolve at late times as stars with progressively longer 
orbital periods reach pericenter and interact with it.
The asymptotic behavior of $s(t)$ in a spherical
non-evolving galaxy in the absence of encounters is roughly
\beq
s(t) = {d\over dt}\left({1\over a}\right) \approx
16\pi^2 K \int f_m(E) dE
\label{eq:drain}
\eeq
where $f_m(E)$ is the phase-space mass density of
stars and $K$ is a constant that defines the mean,
dimensionless change of energy  of the binary in
one interaction with a star \citep{yu-02}.
Equation~(\ref{eq:drain}) equates the rate of change
of the binary's energy with the rate at which stars,
moving along their unperturbed orbits, enter the
binary's influence sphere and extract energy from it.
Depopulation of the orbits as stars are ejected
is represented by
progressively restricting the range of the energy
integral; the lowest allowed energy at any
time $t$ is that
corresponding to an orbit with radial period $P(E)=t$.
Following \cite{yu-02}, we set $K=1.6$, the approximate
value for a ``hard'' binary, and assume that the
integration starts at $a=a_h$.
Figure~\ref{fig:s} compares the solution to Equation~(\ref{eq:drain})
with the $N$-body hardening rates for ($\gamma=0.5,q=0.5$).
Clearly, the ``draining'' of long-period orbits
contributes only minimally to the late evolution seen
in the $N$-body integrations, i.e. the evolution in
phase three is driven almost entirely by loss-cone 
repopulation.

Plots like those in Figure~\ref{fig:s}
were used to estimate $t_{stall}$
for each ($\gamma,M_2/M_1$).
Unavoidably, these estimates are somewhat
subjective.
For the larger values of  $M_2$, 
$t_{stall}$ was taken to be the time when
the $s(t)$ curves for the two different
$N$-values begin to separate.
For the integrations with smaller $M_2$, 
collisional effects are more important, 
and the two $s(t)$ curves separate even
before the peak value is reached.
In these integrations, $t_{stall}$ was 
taken to be the time at which $s(t)$
in the $N=200$K integration reached its
first minimum after the peak.
These estimated $t_{stall}$ values
are likely to be systematically larger than
the true values in a collisionless galaxy.
However the uncertainties in the estimated values 
of $a_{stall}\equiv a(t_{stall})$
and $M_{def}(t_{stall})$ are probably 
modest, at least for the larger $M_2$ values.
For instance, Figure~\ref{fig:binary}
shows that the mass deficit in the $N=200K$
integration of the ($\gamma=1.0,q=0.1$) model
varies only between $0.50\lap M_{def}/M_1\lap 0.65$
for $110\le t\le 140$.
Integrations with larger $N$ will ultimately
be required in order to improve on these estimates.

\begin{figure}
\centering
\includegraphics[scale=0.45,angle=0.]{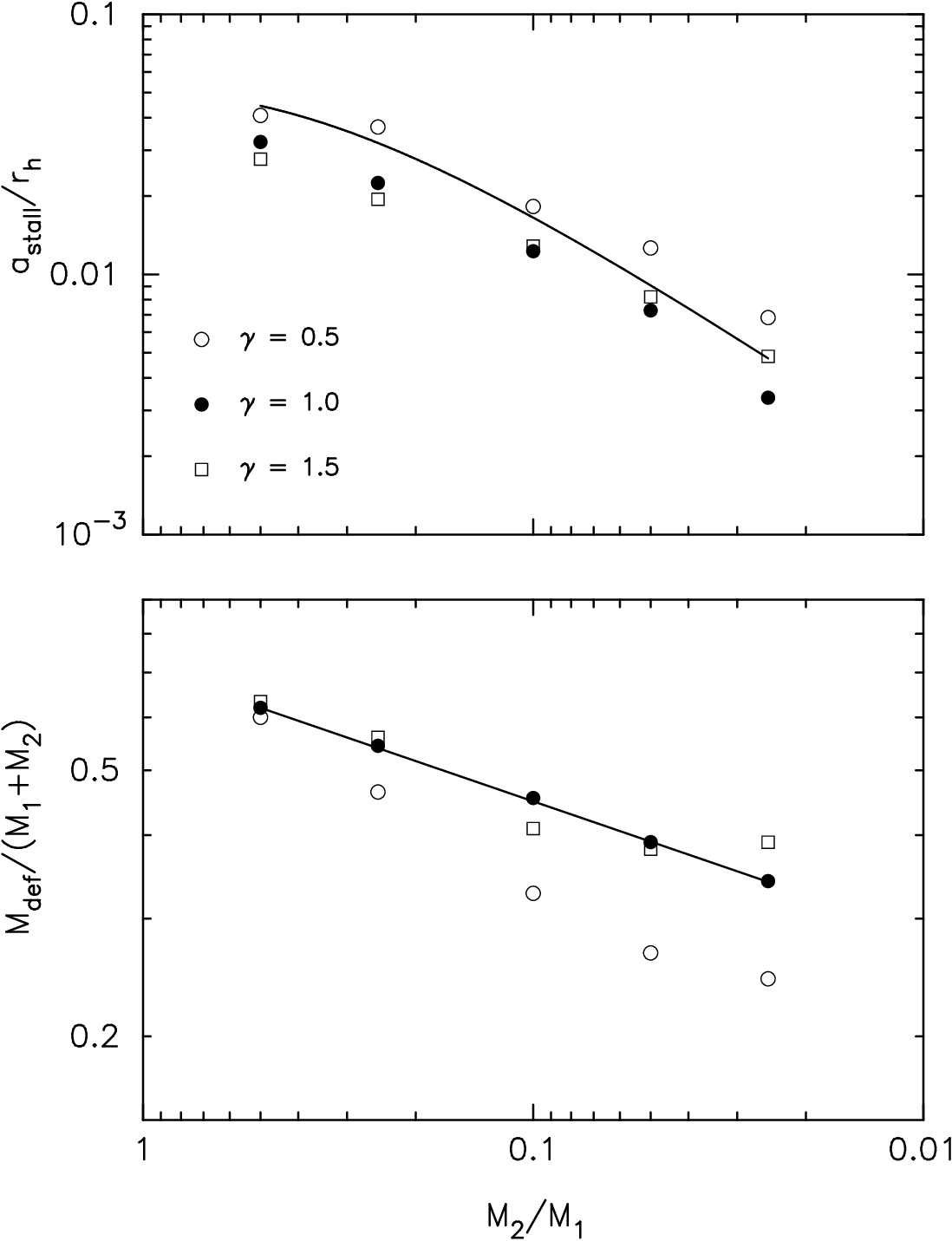}
\caption{Stalling radius (a) and mass deficit at $t=t_{stall}$ (b)
as functions of the binary mass ratio.
Curve in (a) is Equation (12);
curve in (b) is Equation (13).
\label{fig:stall}
}
\end{figure}

The results are presented in Table 1 and 
Figure~\ref{fig:stall}.
The upper panel of Figure~\ref{fig:stall}
shows $a_{stall}$ as a fraction of $r_h'$,
the binary's influence radius (the radius
containing a mass in stars equal to $M_1+M_2$)
at $t_{stall}$.
The measured points are compared with
\beq
{a_{stall}\over r_h'} = 0.2 {q\over (1+q)^2};
\label{eq:ahapproxp}
\eeq
this functional form was motivated by Equation~(\ref{eq:ahapprox}).
Given the likely uncertainties in the estimated
$a_{stall}$ values, Figure~\ref{fig:stall}
suggests that there is no significant dependence
of $a_{stall}/r_h'$ on the initial nuclear
density profile.
This result will be used below to estimate
stalling radii in observed galaxies.

\begin{figure*}
\centering
\includegraphics[scale=0.70,angle=-90.]{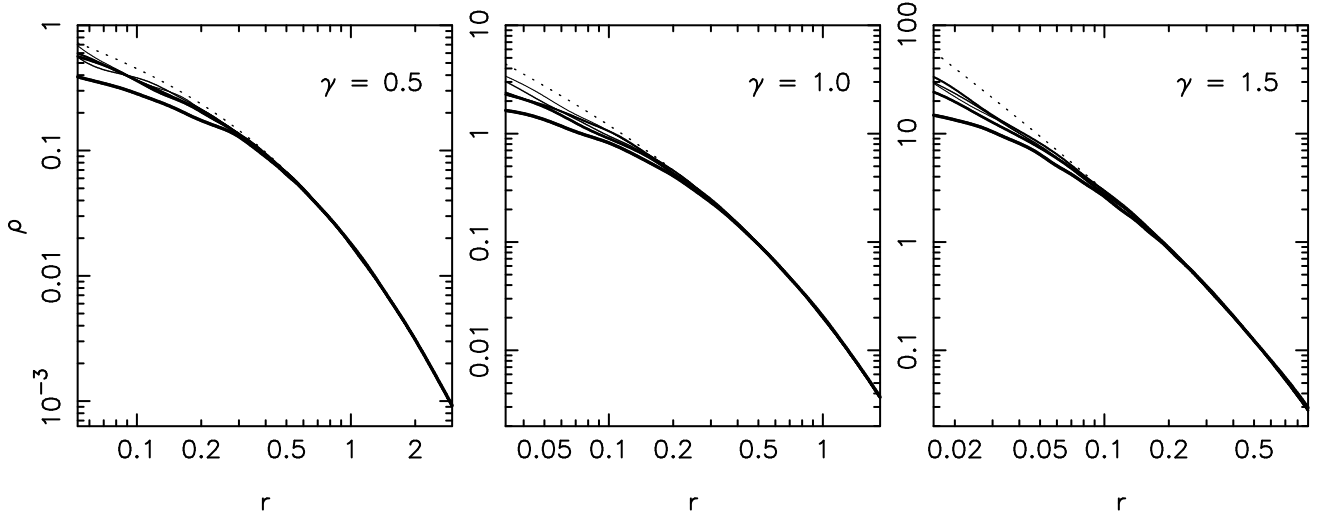}
\caption{Density profiles at $t=t_{stall}$.
Thickness of curves decreases from $q=0.5$ to $q=0.025$.
Dotted lines are the initial density profiles.
\label{fig:profs}
}
\end{figure*}

The lower panel of Figure~\ref{fig:stall}
shows mass deficits at $t=t_{stall}$ as 
a fraction of $M_{12}$.
The line in this figure is
\beq
{M_{def}\over M_{12}} = 0.70 q^{0.2}.
\label{eq:fit}
\eeq
This relation is a good fit to the $\gamma=1.0$ and
$\gamma=1.5$ models, though it overestimates
$M_{def}$ for $\gamma=0.5$.
In any case, the dependence of $M_{def}$ on binary mass ratio
is weak, consistent with the prediction made above.

In galaxies with initially steep nuclear density profiles,
$\rho\sim r^{-\gamma}$, $1.0\lap\gamma\lap 1.5$,
Figure~\ref{fig:stall} suggests that mass
deficits generated by ``stalled'' binaries 
should lie in the narrow range
\beq
0.4\lap {M_{def}\over M_1+M_2} \lap 0.6,\ \ \ \ 
0.05\lap q \lap 0.5.
\eeq
To a good approximation, $M_{def}\approx 0.5 M_\bullet$.

Density profiles at $t=t_{stall}$ are shown in Figure~\ref{fig:profs}.
It is interesting that none of the profiles exhibits the
very flat, nearly constant density cores seen in some
bright elliptical galaxies \citep{kormendy-85,lauer-02,acs-6}.

\section{Multi-stage mergers}

The weak dependence of $M_{def}/M_{12}$ on initial
density profile and on $q$ found above
has an obvious implication:
in repeated mergers, mass deficits should increase
cumulatively,
even if expressed as a multiple of $M_\bullet$, 
the combined mass of the two SBHs at the end of 
each merger.
If the stellar mass displaced in a single
merger is $\sim 0.5 M_{12}$,
then (assuming that the two SBHs always coalesce 
before the next SBH falls in)
the mass deficit following ${\cal N}$ mergers 
with $M_2\ll M_1$ is $\sim 0.5{\cal N}M_\bullet$
with $M_\bullet$ the accumulated mass of the
SBH.

\begin{figure*}
\centering
\includegraphics[scale=0.70,angle=-90.]{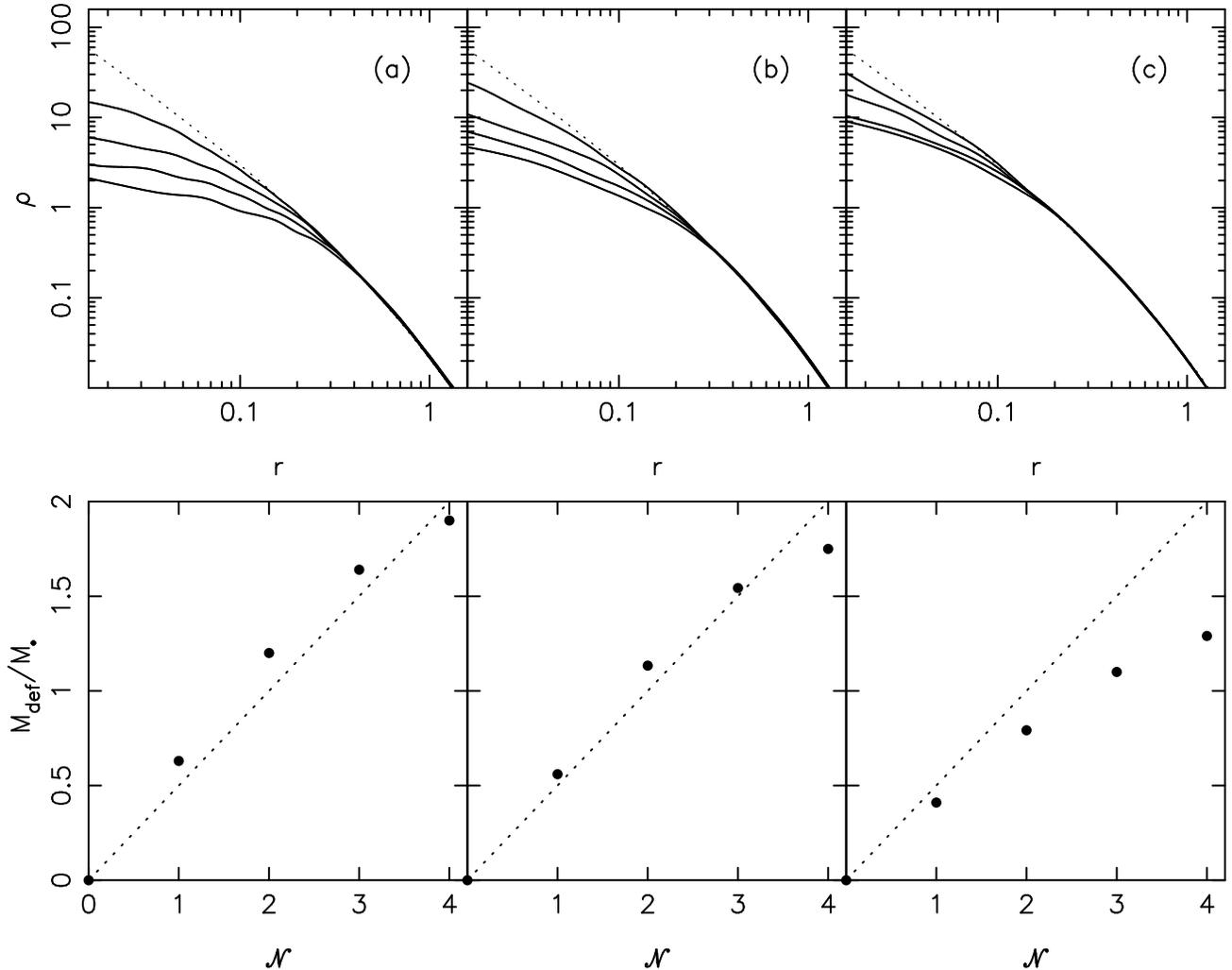}
\caption{Density profiles and mass deficits in the
multi-stage integrations. (a) $M_2=0.005$; (b) $M_2=0.0025$;
(c) $M_2=0.001$.
In the upper panels, dotted lines show the initial 
density profile and solid lines show $\rho(r)$ 
at the end of each merger, i.e. at $t=t_{stall}$.
In the lower panels, 
points show $M_{def}/M_\bullet$ at $t_{stall}$ and
dotted lines show  $M_{def}/M_\bullet=0.5 {\cal N}$,
where $M_\bullet$ is the accumulated SBH mass.
\label{fig:multi}
}
\end{figure*}

Equation~(\ref{eq:fit})  could be used to derive
a more precise prediction for the dependence of
the mass deficit on ${\cal N}$.
But direct simulation is a better approach.
To this end, 
multi-stage $N$-body integrations were carried out.
The starting point for each set of integrations 
was one of the $\gamma=1.5$, $M_1=0.01$  $N$-body models 
described above, extracted at $t=t_{stall}$.
The two massive particles were replaced by a single
particle with mass equal to their combined mass, 
and with position and velocity equal to the center-of-mass
values  for  the binary.
A second massive particle, with mass equal to the
original value of $M_2$, was then added and the model
integrated forward until the new stalling radius
was reached.
The process was then repeated.
In each set of experiments, the mass $M_2$ of the second
SBH particle was kept fixed; in other words,
$M_\bullet$, the accumulated central mass, 
increased linearly with the number ${\cal N}$
of mergers, while the binary mass ratio decreased
with ${\cal N}$.
Three different values of $M_2$ were tried:
$M_2=(0.005,0.0025,0.001)$.

Figure~\ref{fig:multi} and Table 2 give the results.
The cumulative effect of multiple mergers is 
clear from the figure:
$M_{def}/M_\bullet$ increases roughly linearly with
${\cal N}$ for the first few mergers, reaching values
of $\sim 1$ for ${\cal N} = 2$ and
$\sim 1.5$ for ${\cal N} = 3$.
For larger ${\cal N}$ and for smaller $M_2$, 
the increase of $M_{def}/M_\bullet$ with ${\cal N}$
begins to drop below a linear relation.
That this should be so is clear from 
the results of the single-stage mergers
(Figure~\ref{fig:stall}).
Nevertheless, Figure~\ref{fig:multi} verifies
that (at least for $q\gap 0.1$) hierarchical mergers
produce mass deficits that scale roughly
linearly with both ${\cal N}$ and $M_\bullet$
for small ${\cal N}$.

\begin{table}
\begin{center}
\caption{Multi-Stage $N$-body integrations \label{tbl-2}}
\begin{tabular}{cccccc}
\tableline\tableline
$M_2$ & ${\cal N}$ & $M_\bullet$ & $M_{def}$ & $M_{def}/M_\bullet$ & $M_{def}/({\cal N}M_\bullet)$ \\
\tableline
$0.0050$ & 1 & $0.0150$ & $0.0095$ & $0.63$ & $0.63$ \\
         & 2 & $0.0200$ & $0.024$  & $1.20$ & $0.60$ \\
         & 3 & $0.0250$ & $0.041$  & $1.64$ & $0.55$ \\
         & 4 & $0.0300$ & $0.057$  & $1.90$ & $0.48$ \\
&&&&\\
$0.0025$ & 1 & $0.0125$ & $0.0070$ & $0.56$ & $0.56$ \\
         & 2 & $0.0150$ & $0.017$  & $1.13$ & $0.57$ \\
         & 3 & $0.0175$ & $0.027$  & $1.54$ & $0.51$ \\
         & 4 & $0.0200$ & $0.035$  & $1.75$ & $0.44$ \\
&&&&\\
$0.0010$ & 1 & $0.0110$ & $0.004$  & $0.41$ & $0.41$ \\
         & 2 & $0.0120$ & $0.010$  & $0.79$ & $0.40$ \\
         & 3 & $0.0130$ & $0.014$  & $1.10$ & $0.37$ \\
         & 4 & $0.0140$ & $0.018$  & $1.29$ & $0.32$ \\
\tableline
\end{tabular}
\end{center}
\end{table}

Of course, these results 
are only meaningful under the assumption that
the binary manages to coalesce between successive
merger events.
Infall of a SBH into a nucleus containing an
uncoalesced binary would almost certainly result in
{\it larger} values of $M_{def}/M_\bullet$
than those given in Table 2, for reasons discussed below.

\section{Comparison with Observed Mass Deficits}

Mass deficits, computed from observed
luminosity profiles, have been published for a number 
of ``core'' galaxies \citep{mm-02,ravin-02,graham-04,acs-6}.
As first emphasized by \cite{mm-02},
computing $M_{def}$ is problematic due to the unknown
form of the galaxy's luminosity profile before
it was modified by the binary.
\cite{graham-04} noted that Sersic's law provides
the best global fit to the luminosity profiles
of early-type galaxies and bulges and proposed
that mass deficits be defined in terms of the deviation
of the inner profile from the best-fitting Sersic law.
This procedure was followed also by \cite{acs-6}
in their study of Virgo galaxies using HST/ACS data.

\begin{figure*}
\centering
\includegraphics[scale=0.55,angle=-90.]{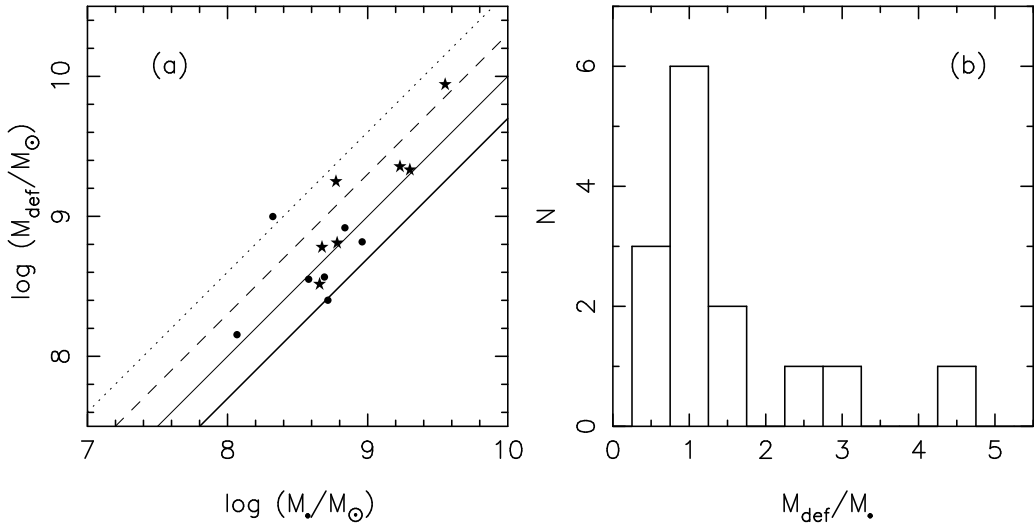}
\caption{(a) Observed mass deficits from Graham (2004)
(filled circles) and Ferrarese et al. (2006) (stars).
Thick, thin, dashed and dotted lines show 
$M_{def}/M_\bullet = 0.5, 1, 2$ and $4$ respectively.
(b) Histogram of $M_{def}/M_\bullet$.
\label{fig:mdefobs}
}
\end{figure*}

Figure~\ref{fig:mdefobs} summarizes the results from 
the \cite{graham-04} and \cite{acs-6} studies.
Mass deficits from \cite{graham-04} were increased
by $0.072$ in the logarithm to correct
an error in the mass-to-light ratios (A. Graham,
private communication).
\cite{graham-04} gives two estimates of $M_\bullet$
for each galaxy, based on the empirical correlation
with concentration parameter \citep{graham-01}
and on the \cite{gebhardt-00} version of the
$M_\bullet-\sigma$ relation.
We recomputed SBH masses for Graham's galaxies
using the most current version of the $M_\bullet-\sigma$
relation \citep{ff-05}.
\cite{acs-6} computed $M_\bullet$ for their sample 
galaxies in the same way;
we have replaced $M_\bullet$ for three of their
galaxies (NGC4486, NGC4374, NGC4649) with 
values derived from detailed kinematical modelling
\citep{macchetto-97,bower-98,gebhardt-03}.
One ``core'' galaxy from \cite{acs-6} -- NGC 798 -- 
was excluded since it contains a large-scale stellar disk.

The  histogram of $M_{def}/M_\bullet$ values is shown in
Figure~\ref{fig:mdefobs}b.
There is a clear peak at $M_{def}/M_\bullet\approx 1$.

Sersic's law -- which describes the projected, or surface,
luminosity density of early-type galaxies -- implies a {\it space} 
density that varies
as $j(r) \sim r^{-\gamma}$ at small radii, where
$\gamma\approx (n-1)/n$ and $n$ is the Sersic index.
The galaxies plotted in Figure~\ref{fig:mdefobs} have
$n \gap 5$, hence $0.8\lap \gamma\lap 1.0$; in other words,
the mass deficits for these galaxies were computed under the assumption
that the pre-existing nuclear density was a power law
with $\gamma\approx 1$.
This is approximately the same inner dependence as in 
our $\gamma=1$ Dehnen-model galaxies.
While the cumulative $N$-body mass deficits shown in Figure~\ref{fig:multi}
and Table~2 were based on initial models with $\gamma=1.5$,
Figure~\ref{fig:stall} suggests that the results would
have been almost identical for
$\gamma=1$, at least during the first few stages of the
merger hierarchy.

Comparing Figures~\ref{fig:multi} and \ref{fig:mdefobs},
we therefore conclude that most ``core''
galaxies have experienced $1\lap {\cal N} \lap 3$ mergers
(i.e. $0.5\lap M_{def}/M_\bullet \lap 1.5$),
with ${\cal N}=2$ the most common value.

A few galaxies in Figure~\ref{fig:mdefobs}
have significantly larger mass deficits; 
the most extreme object is NGC 5903 with
$M_{def}/M_\bullet\approx 4.5$.
Of course the uncertaintities associated
with the $M_{def}$ and $M_\bullet$ values
in Figure~\ref{fig:mdefobs} may be large; the former due
to uncertain $M/L$ corrections, the latter due 
to various difficulties associated with SBH
mass estimation (e.g. Maciejewski \& Binney 2001;
Valluri, Merritt \& Emsellem 2004).
In the case of NGC 5903, for which $M_\bullet$
was computed from the $M_\bullet-\sigma$ relation,
we note that published velocity dispersions for
this galaxy range from $192$ km s$^{-1}$ \citep{smith-00}
to $245$ km s$^{-1}$ \citep{davies-87};
the corresponding range in the inferred 
black hole mass is 
$1.36\times 10^8 \lap M_\bullet/M_\odot \lap 4.45\times 10^8$
and the range in $M_{def}/M_\bullet$ is 
$1.9 \lap M_{def}/M_\bullet \lap 6.2$.
Even if $\sigma$ were known precisely,
there will always be a substantial uncertainty
associated with any $M_\bullet$ value derived from 
an empirical scaling relation.

But assuming for the moment that the numbers plotted
in Figure~\ref{fig:mdefobs} are accurate,
how might the galaxies with $M_{def}/M_\bullet > 2$ be explained?
There are several possibilities.
(a) These galaxies are the product of ${\cal N}\gap 4$ mergers.
Two of the largest $M_{def}/M_\bullet$ values 
in Figure~\ref{fig:mdefobs}
are associated with M87 and M49, both extremely luminous
galaxies that could have experienced multiple mergers.
However, Figure~\ref{fig:multi} suggests that
$M_{def}/M_\bullet$ values $\gap 3$
might be difficult to produce via repeated mergers.
(b) The primordial density profiles in  these galaxies
were flatter than Sersic's law at small radii.
(c) The two SBHs failed to efficiently coalesce
during one or more of the merger events,
so that a binary was present  
when a third SBH fell in.
Indeed this is likely to be the case
in the largest (lowest density)
galaxies in Figure~\ref{fig:mdefobs},
which have little gas and extremely long
timescales for loss-cone repopulation
by two-body relaxation.
Infall of a third SBH onto an uncoalesced binary
is conducive to smaller values
of $M_\bullet$, since one or more of the SBHs
could eventually be ejected by the gravitational
slingshot \citep{mv-90}; and to larger values  of $M_{def}$,
since multiple SBHs are more efficient than a binary
at displacing stars \citep{merritt-04b}.
(The former point is only relevant to
galaxies -- unlike NGC 5903 -- for which $M_\bullet$ 
has actually been measured.)
The chaotic interaction between three SBHs would
probably also assist in the gravitational wave
coalescence of the two most massive holes by inducing
random changes in their relative orbit \citep{blaes-02}.
(d) The gravitational-wave rocket effect
is believed capable of delivering kicks
to a coalescing binary as large as $\sim 200$ km s$^{-1}$
\citep{favata-04,blanchet-05,herrmann-06,baker-06}.
The stellar density drops impulsively when the SBH 
is kicked out, and again when its orbit
decays via dynamical friction.
Mass deficits produced in this way can be as large as 
$\sim M_\bullet$ \citep{merritt-04a,copycats-04}.
(e) Stars bound to the infalling SBH -- which were
neglected in the $N$-body simulations presented here --
might affect $M_{def}$, although it is not clear
what the direction or magnitude of the change would be.

Focussing again on the majority of galaxies in 
Figure~\ref{fig:mdefobs} with $0.5\lap M_{def}/M_\bullet \lap 1.5$, 
we can ask whether values of ${\cal N}$ in the range
$1\le {\cal N} \le 3$ are consistent with
hierarchical models of galaxy formation.
If the seeds of the current SBHs were present at
large redshift, 
the ancestry of a bright galaxy could include dozens
of mergers involving binary SBHs \citep{volon-03a,sesana-04}.
However the more relevant quantity is probably the number
of mergers since the era at  which most of the gas
was depleted, since star formation from ambient gas
could re-generate a density cusp after its destruction by 
a binary SBH \citep{graham-04}.
\cite{hk-02} calculate just this quantity,
based on semi-analytic models for galaxy mergers that
include prescriptions for star- and SBH-formation.
Their Figure~2 shows probability distributions for
${\cal N}$ as a function of galaxy luminosity for
mergers with $q>0.3$.
For galaxies like those in Figure~\ref{fig:mdefobs}
($M_V\lap -21, M_B\lap -20$),
\cite{hk-02} find a median ${\cal N}$ of $\sim 1$,
but particularly among the brightest galaxies,
they find that values of ${\cal N}$ as large as 3 or 4 
are also likely.
This prediction is consistent with Figure~\ref{fig:mdefobs}.

Probably the biggest uncertainty in this analysis
is the unknown behavior of the binary after it 
reaches $a\approx a_{stall}$
and before the next SBH falls in.
If mergers are ``dry,''
gas-dynamical torques can not be invoked to
accelerate the coalescence, implying a complicated
interaction between three SBHs when the next merger event occurs.
As discussed above, the net effect would be an
increase in $M_{def}/M_\bullet$.
On the other hand, if gas {\it is} present in
sufficient quantities to assist in coalescence,
it may also form new stars -- decreasing
$M_{def}$ -- and/or accrete onto the SBH -- increasing
$M_\bullet$; in either case, $M_{def}/M_\bullet$ would 
be smaller than computed here. 

\section {Stalled Binary Separations}

Equation~(\ref{eq:ahapproxp}) gives an estimate
for the stalling radius of a binary SBH in terms 
of its influence radius $r_h'$; the latter is defined
as the radius of a sphere containing a stellar mass 
equal to twice $M_\bullet\equiv M_1+M_2$, {\it after}
the binary has reached $a_{stall}$, i.e. after it has
finished modifying the stellar density profile.
One can use Equation~(\ref{eq:ahapproxp})
to estimate binary separations
in galaxies where $M_\bullet$ and $\rho(r)$ are known,
under the assumption (discussed in more detail below)
that no additional mechanism has induced 
the binary to evolve beyond $a_{stall}$.
Table 3 gives the results for the seven Virgo
``core'' galaxies shown in Figure~\ref{fig:mdefobs}.
Influence radii were computed using
parametric (``core-Sersic'') fits to the luminosity profiles,
as described in \cite{acs-6}; that paper also gives the
algorithm which was used for converting luminosity densities
into mass densities.
When discussing real galaxies, $r_h'$ is just the
currently-observed influence radius $r_h$ and 
henceforth the prime is dropped.

Stalling radii are given in Table~3 
assuming $q=0.5$ and $q=0.1$; $a_{stall}$ for other values
of $q$ can be computed from Equation~(\ref{eq:ahapproxp}).
Table 3 also gives angular sizes of the binaries
assuming a distance to Virgo of 16.52 Mpc.
Typical separations between components in a stalled binary
are found to be $\sim 10\ {\rm pc}\ (0.1'')$ ($q=0.5$) and 
$\sim 3\ {\rm pc}\ (0.03'')$ ($q=0.1$).
Of course these numbers should be interpreted as upper limits;
nevertheless they are large enough to suggest that binary
SBHs might be resolvable in the brighter Virgo galaxies
if both SBHs are luminous.

\begin{table}
\begin{center}
\caption{Virgo ``Core'' Galaxies \label{tbl-3}}
\begin{tabular}{ccccccccc}
\tableline\tableline
       &       &             &        & $a_{\rm stall}$ & $a_{stall}$ &          & $v_{esc}$ & $v_{esc}$ \\
Galaxy & $B_T$ & $M_\bullet$ & $r_h'$ & $q=0.5$         & $q=0.1$     & $v_{ej}$ & $q=0.5$   & $q=0.1$   \\
(1) & (2) & (3) & (4) & (5) & (6) & (7) & (8) & (9) \\
\tableline
NGC 4472 & -21.8 & $5.94$ & $130.\ (1.6) $ & $5.6\ (0.070)$ & $2.1\ (0.026)$ & 562. & $1395.$ & $1865.$ \\
NGC 4486 & -21.5 & $35.7$ & $460.\ (5.7) $ & $20.\ (0.25)$  & $7.6\ (0.095)$ & 733. & $1480.$ & $2175.$ \\
NGC 4649 & -21.3 & $20.0$ & $230.\ (2.9) $ & $10.\ (0.13)$  & $3.8\ (0.047)$ & 776. & $1590.$ & $2325.$ \\
NGC 4406 & -21.0 & $4.54$ & $90.\  (1.1) $ & $4.0\ (0.050)$ & $1.5\ (0.019)$ & 590. & $1255.$ & $1790.$ \\
NGC 4374 & -20.8 & $17.0$ & $170.\ (2.1) $ & $7.6\ (0.094)$ & $2.8\ (0.035)$ & 832. & $1635.$ & $2435.$ \\
NGC 4365 & -20.6 & $4.72$ & $115.\ (1.4) $ & $5.0\ (0.063)$ & $1.9\ (0.023)$ & 533. & $1115.$ & $1615.$ \\
NGC 4552 & -20.3 & $6.05$ & $73.\  (0.91)$ & $3.2\ (0.040)$ & $1.2\ (0.015)$ & 757. & $1500.$ & $2230.$ \\
\tableline
\end{tabular}
\end{center}
Notes. -- 
Col. (1): New General Catalog (NGC) numbers.
Col. (2): Absolute $B$-band galaxy magnitudes.
Col. (3): Black hole masses in $10^8M_\odot$.
Col. (4): Black hole influence radii, defined as the radii containing a mass 
in stars equal to $2M_\bullet$, in pc (arcsec).
Col. (5): Binary stalling radii for $q=0.5$, in pc (arcsec).
Col. (6): Binary stalling radii for $q=0.1$, in pc (arcsec).
Col. (7): Typical ejection velocity from a binary SBH with $a=a_{stall}$ (km s$^{-1}$).
Cols. (8), (9): Escape velocity (km s$^{-1}$).
\end{table}

\bigskip\bigskip

\section{Discussion}

We have shown that
the mass deficit produced by a binary SBH at
the center of a spherical galaxy is ``quantized'' in units of 
$\sim 0.5(M_1+M_2)$, with only
a weak dependence on binary mass ratio, and
that $M_{def}/M_\bullet$ grows approximately linearly
with  the number of merger events.
These results were compared with observed mass deficits 
to conclude that most bright elliptical galaxies have 
experienced $1-3$ mergers since the era at which
gas was depleted and/or star formation became inefficient.

In this section, we explore some further implications 
of the $N$-body results, and discuss their generality.

\subsection{Loss-Cone Refilling and Non-Spherical Geometries}

The approach adopted in this paper was motivated 
by the fact that mass deficits, or cores, are only observed
at the centers of galaxies with very long
relaxation times.
In such galaxies, collisional repopulation of orbits
depleted by the binary would act too slowly to
significantly influence the binary's evolution \citep{valtonen-96},
and the binary would stop evolving
once it had interacted with all stars on intersecting orbits.
But even in the completely collisionless case,
evolution of the binary, and its effect on the local
distribution of stars, could be different in nonspherical
(axisymmetric, triaxial, ...) geometries.
There are really two questions here: 
(a) How does $a_{stall}$ depend on geometry?
(b) If a binary continues to harden below $a_{stall}$,
what would be the effect on $M_{def}$?

The answer to the latter question is complex.
Consider a galaxy containing a stalled binary, and
suppose that some mechanism -- gravitational encounters,
time-dependent torques from a passing galaxy,
perturbations from a third SBH, new star formation, etc. 
-- has the net effect
of placing additional stars on orbits that intersect
the binary.
These stars will be ejected, and the binary will shrink.
However there need not be any net change in the mass deficit,
since the new stars are first added, then subtracted,
from the nuclear density.
(Second-order changes in $\rho$ due to the time dependence of the 
gravitational potential are being ignored.)
This argument suggests that the good correlation found here
(Figure~\ref{fig:stall}) between 
$a_{stall}$ and $M_{def}$ need not apply more generally,
in  time-dependent or non-spherical situations.

An example of the latter is a binary embedded in a triaxial
galaxy which contains centrophilic (box or chaotic) orbits.
The mass in stars on centrophilic orbits
can be extremely large, $\gg M_\bullet$, in  a triaxial galaxy,
although many orbital periods will generally be required
before any given star passes near enough to the binary to 
interact with it \citep{poon-04,hb-06}.
Nevertheless the binary need not stall \citep{bmsb-06}.
Self-consistent calculations of the effect of a binary
on the nuclear density profile have not been carried
out in the triaxial geometry, but it is clear that mass deficits
might be smaller than in the spherical geometry since
some stars ejected by the binary will be on orbits
with very large ($\gg r_h$) characteristic radii.

The influence of different geometries on $a_{stall}$
and $M_{def}$ will be investigated in future
papers.

\subsection{Mass Deficits and Core Radio Power}

An intriguing relation was discovered by \cite{bc-06}
and \cite{cb-06}
between the radio and morphological properties of active
galaxies.
Radio-loud AGN -- active nuclei with a large ratio of radio
to optical or radio to X-ray luminosities --
are {\it uniquely} associated with ``core'' galaxies.
In fact, the slope of the inner brightness profile appears to be
the {\it only} quantity  that reliably predicts whether an AGN
is radio-loud or radio-quiet.
\cite{cb-06} also showed that when the radio-loud
and radio-quiet galaxies are considered separately, there is no
dependence of radio loudness on the nuclear density profile 
within either class.
In other words, the fact that an active galaxy is morphologically
a ``core'' galaxy predicts that it will be radio-loud but
does not predict how loud.
While the origin of this connection is unclear,
it might reflect the influence of SBH rotation on radio
power \citep{wc-95}: 
``core'' galaxies have experienced a recent merger,
and coalescence of the two SBHs resulted in a rapidly-spinning
remnant.

Two results from the $N$-body work presented here may
be relevant to the Capetti \& Balmaverde correlation.
First, as shown above, the mass deficit is a weak function 
of the mass ratio of the binary that produced it, i.e., 
the shape of the nuclear density profile
is poorly correlated with the mass of the infalling
hole.
Since the degree of spin-up is also weakly correlated
with $M_2/M_1$ \citep{wc-95,me-02,hb-03} -- even
very small infalling holes can spin up the larger
hole to near-maximal spins --
it follows that SBH rotation should be poorly correlated
with mass deficit.
This may explain the weak dependence of radio
loudness on profile slope found by \cite{cb-06}
for the core galaxies.

Second, as argued above, core galaxies may contain
uncoalesced binaries.  Perhaps the presence of a second
SBH is the main factor that determines radio loudness.

\subsection{High-Velocity Stars}

Gravitational slingshot ejections by a binary SBH 
produce a population
of high-velocity stars with trajectories directed
away from the center of the galaxy.
A few hyper-velocity stars have been detected in the 
halo of the Milky Way \citep{brown-05,hirsch-05,brown-06},
and a binary SBH at the Galactic center is a possible
model for their origin.
The short nuclear relaxation time would accelerate
the hardening of a massive binary by ensuring that stars were
scattered into its sphere of influence \citep{yu-03}.
In this model, stars could have been ejected 
during the late stages of the binary's evolution,
when $a\ll a_h$,
at high enough velocities and large enough numbers
to explain the observed hyper-velocity stars.

Binary SBHs have also been invoked to explain other
populations, e.g. intergalactic planetary nebulae
in  the Virgo cluster \citep{arna-04}.
If every Virgo galaxy harbored a binary SBH with mass
ratio $q=0.1$ which evolved, via slingshot 
ejection of stars, to the gravitational-radiation regime,
the total mass ejected would have been $\sim 2\%$
of the luminous mass of the cluster \citep{hb-05}.
But at least in the brighter Virgo galaxies -- including
the galaxies in Table 3 --
two-body relaxation times are much too long for
collisional resupply of binary loss cones, and there
is no compelling reason to assume that the binary SBHs in these 
galaxies would have continued interacting with stars 
beyond $a_{stall}$.
At these separations, velocities between 
the two components of the binary are relatively modest,
implying a much lower probability of ejection of stars 
at velocities large enough to escape the galaxy.

The mean specific energy change of a star that interacts with
a hard binary is $\sim 3G\mu/2a$ \citep{hills-83,quinlan-96}.
Combining this with Equation~(\ref{eq:ahapproxp}) for $a_{stall}$
gives the typical ejection velocity from a stalled binary:
\beq
{\rm v}_{ej}\approx 4.0 \left({GM_{12}\over r_h}\right)^{1/2}
\approx 830\ {\rm km} {\rm s}^{-1} 
\left({M_{12}\over 10^8M_\odot}\right)^{1/2}
\left({r_h\over 10\ {\rm pc}}\right)^{-1/2},
\label{eq:vej}
\eeq
independent of mass ratio.
Of course this is an upper limit in the sense
that ejection velocities are 
lower when $t<t_{stall}$.
Table 3 gives ${\rm v}_{ej}$ for the Virgo ``core'' galaxies
and also ${\rm v}_{esc}$, the escape velocity, 
defined as $\sqrt{-2\Phi(a_{stall})}$
with $\Phi(r)$ the gravitational potential
including the contribution from the binary,
modelled as a point with mass equal to the currently-observed
value of $M_\bullet$.
Ejection velocities are seen to be much smaller than 
${\rm v}_{esc}$ in all cases, implying that essentially no stars would
be ejected into the intracluster medium.

\subsection {Dark-Matter Cores}

The specific energy change of a dark matter particle
interacting with a binary SBH is identical to that of a star.
At least in the case of bright elliptical galaxies like those 
in Table 3, it is unlikely that dark matter was ever a dominant
component near the center or that it significantly influenced
the evolution of the binary.
But ejection of dark matter particles by a massive binary
would produce a core in the dark matter distribution similar 
in size to the luminous-matter core, $r_{core}\approx r_h$.
A more definite statement about the variation with radius of
$\rho_{DM}/\rho_\star$ near the center of a galaxy containing
a binary SBH is probably impossible to make
without $N$-body simulations that contain all three components.
However Table 3 suggests dark matter core radii of
hundreds of parsecs in bright elliptical galaxies.

It follows that the rates of self-interaction
of supersymmetric particles at the centers of 
galaxies like M87 \citep{baltz-00} would be much lower than 
computed under the assumption that the dark matter 
still retains its ``primordial'' density profile 
\citep{NFW-96,moore-98}.
This fact has implications for so-called ``indirect'' dark matter 
searches, in which inferences are drawn about the
properties of particle dark matter based on measurements
of its self-annihilation by-products \citep{bhs-04}.
One of the proposed search strategies is to identify
a component of the diffuse gamma-ray background that
is generated by dark matter annhilations
in halos at all redshifts \citep{ullio-02,ts-03}.
Calculations of the background flux \citep{ando-05,em-05}
have so far always assumed that the dark matter distribution
does not change with time.
Since the annihilation flux in a smooth dark matter halo
is dominated by the the center, 
these calculations may substantially over-estimate the contribution 
of galactic halos
to the gamma-ray background.

\acknowledgments
I thank Pat Cote, Laura Ferrarese and Alister Graham for
help with the galaxy data that were analyzed in \S6
and for illuminating discussions.
This work was supported by grants 
AST-0071099, AST-0206031, AST-0420920 and AST-0437519 from the 
NSF, grant NNG04GJ48G from NASA,
and grant HST-AR-09519.01-A from STScI.
The $N$-body calculations presented here were carried out at
the Center for the Advancement of the Study of 
Cyberinfrastructure at RIT whose support is gratefully
acknowledged.

\end{document}